\documentclass{article}

\usepackage{arxiv}

\usepackage[utf8]{inputenc} 
\usepackage[T1]{fontenc}    
\usepackage{url}            
\usepackage{booktabs}       
\usepackage{amsfonts}       
\usepackage{nicefrac}       
\usepackage{microtype}      
\usepackage{lipsum}
\usepackage{graphicx}
\usepackage{orcidlink}
\usepackage{comment}
\usepackage{amsmath}
\usepackage{natbib}
\usepackage{tikz}
\usepackage{tcolorbox}
\usepackage{rotating}
\usetikzlibrary{positioning, shapes, arrows.meta}
\definecolor{VUB_blauw}{rgb}{0.1529, 0.2667, 0.5529}
\definecolor{mygrey}{HTML}{949494}
\usepackage{hyperref}       
\hypersetup{
    colorlinks,%
    citecolor=VUB_blauw,%
    filecolor=VUB_blauw,%
    linkcolor=VUB_blauw,%
    urlcolor=VUB_blauw
}
\graphicspath{ {./images/} }
\newcommand{\customCor}[1]{%
  \includegraphics[height=1em]{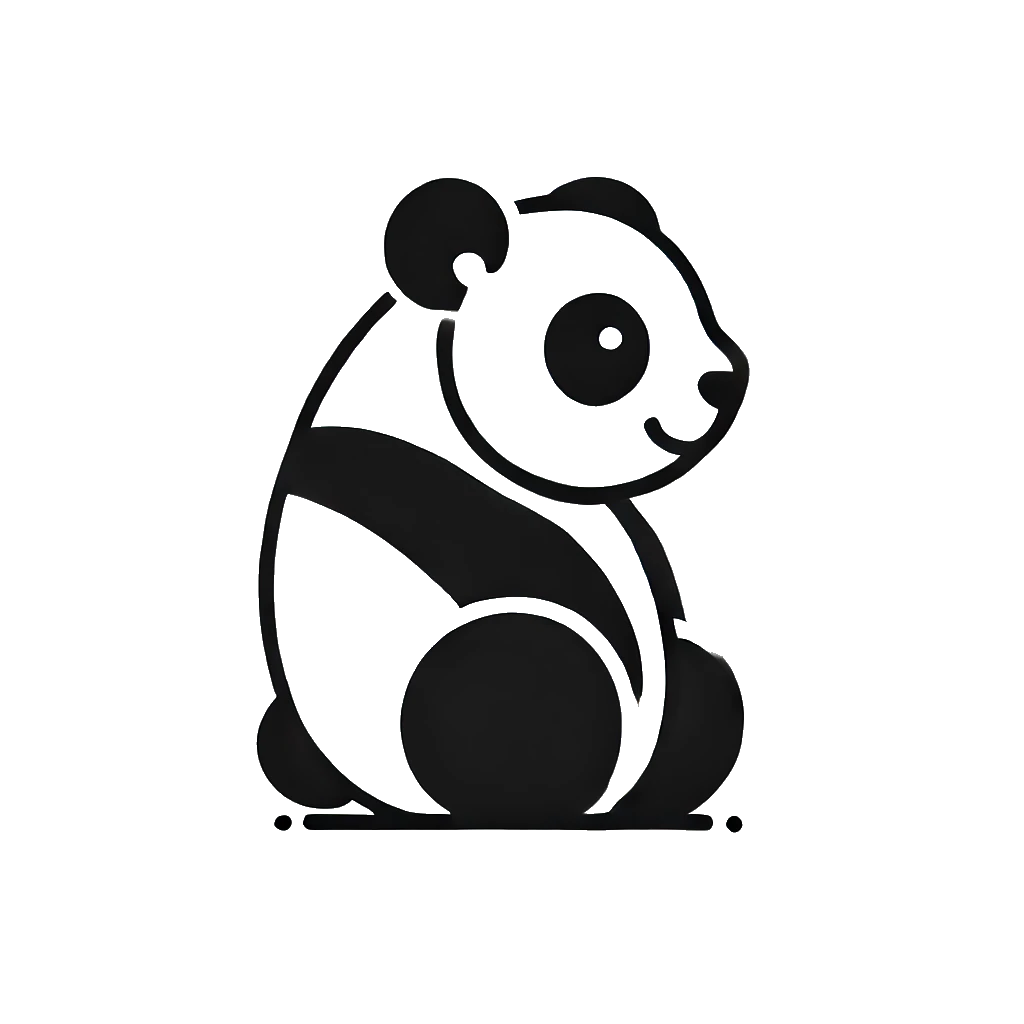} #1%
}

\AddToHook{shipout/background}{%
  \ifnum\value{page}=1 
    \pagenumbering{Roman} 
    \setcounter{page}{1} 
  \fi
  \ifnum\value{page}=2 
  \pagenumbering{arabic} 
  \setcounter{page}{2} 
  \fi
}

\title{Human-in-the-Loop LLM Grading for Handwritten Mathematics Assessments}

\runningtitle{GPT-5.1 Performing Human-Aligned Grading in Handwritten Math Tests}

\author{
  Arne Vanhoyweghen\textsuperscript{1,\customCor{ }} \\ 
  \orcidlinkc{0000-0003-0103-4715} \\
  \And
  Vincent Holst \textsuperscript{1} \\ 
  \orcidlinkc{0009-0002-4117-4966} \\
  \And
  Melika Mobini \textsuperscript{1} \\ 
  \orcidlinkc{0009-0006-2335-5869} \\
  \And
  Lukas Van de Voorde \textsuperscript{1} \\ 
  \orcidlinkc{0009-0006-8797-0676} \\
  \And
  Tibo Vanleke \textsuperscript{1} \\ 
  \orcidlinkc{0009-0003-2200-7529} \\
  \And
  Bert Verbruggen \textsuperscript{1} \\ 
  \orcidlinkc{0000-0001-9776-2420} \\
  \And
  Brecht Verbeken \textsuperscript{1} \\ 
  \orcidlinkc{0000-0002-7506-3298} \\
  \And
  Andres Algaba \textsuperscript{1} \\ 
  \orcidlinkc{0000-0002-0532-3066} \\
  \And
  Sam Verboven \textsuperscript{1} \\ 
  \orcidlinkc{0000-0002-1742-5561} \\
  \And
  Marie-Anne Guerry \textsuperscript{1} \\ 
  \orcidlinkc{0000-0001-5842-8905} \\
  \And
  Filip Van Droogenbroeck \textsuperscript{1} \\ 
  \orcidlinkc{0000-0003-1133-3495} \\
  \And
  Vincent Ginis \textsuperscript{1,2} \\ 
  \orcidlinkc{0000-0003-0063-9608} \\
  \and
  \textsuperscript{1}Data Analytics Lab, Vrije Universiteit Brussel, 1050 Brussel, Belgium \\ 
  \textsuperscript{2}School of Engineering and Applied Sciences, Harvard University, Cambridge, Massachusetts 02138, USA
}
    
\begin{document}
\maketitle
\renewcommand{\thefootnote}{} 
\footnotetext{\includegraphics[height=1em]{panda2.png} Corresponding author: arne.vanhoyweghen@vub.be}
\renewcommand{\thefootnote}{\arabic{footnote}} 
\thispagestyle{plain} 

\begin{abstract}

Providing timely and individualised feedback on handwritten student work is highly beneficial for learning but difficult to achieve at scale. This challenge has become more pressing as generative AI undermines the reliability of take-home assessments, shifting emphasis toward supervised, in-class evaluation. We present a scalable, end-to-end workflow for LLM-assisted grading of short, pen-and-paper assessments. The workflow spans (1) constructing solution keys, (2) developing detailed rubric-style grading keys used to guide the LLM, and (3) a grading procedure that combines automated scanning and anonymisation, multi-pass LLM scoring, automated consistency checks, and mandatory human verification. We deploy the system in two undergraduate mathematics courses using six low-stakes in-class tests. Empirically, LLM assistance reduces grading time by approximately 23\% while achieving agreement comparable to, and in several cases tighter than, fully manual grading. Occasional model errors occur but are effectively contained by the hybrid design. Overall, our results show that carefully embedded human-in-the-loop LLM grading can substantially reduce workload while maintaining fairness and accuracy.

\end{abstract}

\section{Introduction}

Providing timely, consistent, and individualised feedback on student work is widely recognised as one of the most effective ways to support learning and improve motivation across educational contexts~\cite{hattie1999influences,fisher2025impact, young2021tired, shute2008focus}. Research has shown that frequent, low-stakes assessment improves student understanding and motivation~\cite{bangert1991effects, black2009developing, leeming2002exam}, and that active, feedback-rich learning environments yield significant performance gains~\cite{freeman2014active}. Crucially, the benefits of both formative assessment and active learning depend on tight feedback loops: students must receive corrections and explanations quickly enough for those insights to influence subsequent learning~\cite{black2009developing, shute2008focus}. 
Parallel evidence demonstrates the large impact of individualised feedback \cite{schwarcz2017impact} and even greater gains from one to one tutoring \cite{bloom1984two, elbaum2000effective}. Despite this strong pedagogical foundation, providing frequent, high-quality, tailored feedback at scale is difficult, particularly for non-digital or open-ended forms of assessments~\cite{livingston2009constructed}. For instance, handwritten work is especially time-intensive to grade and digitise. These constraints make it challenging for instructors to provide personalised and timely feedback, especially in settings with heavy teaching loads or growing class sizes.

A natural response is to rely more heavily on multiple-choice (MC) formats, which offer administrative efficiency and instant grading. Well-designed MC items effectively assess conceptual understanding, factual knowledge, and broad curriculum coverage, with established best practices for mitigating common pitfalls~\cite{butler2018multiple}. However, MC formats provide limited direct evidence of solution processes, intermediate reasoning, and diagnostic misconceptions—particularly critical when assessment goals emphasise method and partial credit allocation~\cite{livingston2009constructed}. Moreover, MC items risk undersampling valid solution paths and conflating reasoning with test-taking strategies or recognition~\cite{roediger2005positive, butler2018multiple, ozuru2013comparing, melovitz2018analysis}. Short open-ended handwritten responses therefore remain essential for evaluating generative reasoning and structured problem solving, despite substantial operational costs in grading time and feedback timeliness.

Within the modern classroom, large language models introduce both a potential response to this challenge and a new source of tension~\cite{lehmann2024ai, bastani2024generative, miller2024ai, putri2025chalkboards}. On the one hand, recent studies have shown that LLMs can support the grading of open-ended mathematics responses by providing useful baseline scores when guided by structured rubrics and solutions~\cite{gandolfi2025gpt}, and that multimodal LLMs can extend this support to handwritten responses~\cite{liu2024ai}, thereby reducing grading effort and supporting consistency across evaluators. These findings suggest that LLMs can play a meaningful role as assistive tools in assessment workflows, particularly when human verification remains in place. However, existing studies primarily demonstrate grading feasibility or accuracy in controlled settings, leaving open how such systems can be deployed end-to-end in real classrooms, under constraints of privacy, model stochasticity, and repeated use.

On the other hand, the same models increasingly undermine many traditional forms of assessment. Commercial systems~\cite{nano_banana_pro} can now generate high-quality handwritten mathematical solutions, complete with intermediate steps and correct formatting, within seconds. As a result, take-home assignments are becoming less reliable indicators of students' independent understanding. At the same time, as students gain access to increasingly capable AI-based tutoring systems outside the classroom~\cite{henkel2024effective, putri2025chalkboards}, the responsibility shifts further toward educational institutions to provide trustworthy, in-person demonstrations of mastery and to deliver clear and timely signals of student progress. Short, supervised, pen-and-paper assessments are one such response: they reliably capture independent reasoning while remaining compatible with formative assessment practices. Yet these assessments reintroduce substantial workload, as handwritten work must be graded, digitised, and returned with prompt, actionable feedback. For such feedback to be pedagogically effective, it must address not only final answers but also the intermediate reasoning steps students employ. It is precisely in this context that LLMs, which can perform both optical character recognition (OCR) and contextual analysis, can be reframed as part of the solution to the assessment challenges they themselves have exacerbated.

In this manuscript, we present a complete, scalable workflow that integrates LLMs into the grading and assessment cycle for short, supervised, pen-and-paper assessments. Moreover, we assess whether LLM-assigned scores fall within the range of disagreement observed between human graders, using inter-annotator differences as a benchmark for acceptable grading variation. The workflow was deployed in two undergraduate mathematics courses at the Vrije Universiteit Brussel, where students completed a series of brief in-class ``bonus tests'' consisting of two handwritten exercises each. These tests are designed as low-stakes, formative checkpoints: they promote regular engagement, provide students with frequent snapshots of their progress, and give instructors timely insight into emerging difficulties. Because such assessments are repeated and must be returned promptly to be pedagogically effective, they provide an ideal setting to examine how LLMs can support efficient and reliable grading. Our workflow preserves student privacy by pseudonymising submissions internally (mapping student identities to an internal code) and sending the LLM only cropped answer-box images from which all identifiers have been removed. To improve robustness, each response is evaluated independently five times by the LLM. We then run automated consistency checks across the five outputs and conclude with a final human verification step before releasing grades.

We show that, once operationalised, using LLMs as a structured grading baseline reduces grading time by approximately 16\% while yielding grading alignment comparable to, or tighter than, fully manual grading. This estimate errs on the conservative side, since our manual condition was already streamlined (grades entered into an ordered spreadsheet with pre-linked student identifiers, avoiding separate digitisation and ID lookups). These results indicate that LLMs can support timely and consistent grading at scale, provided that human verification remains in place for rare outlier cases. 

The full implementation of our LLM-assisted grading pipeline, including exam sheet processing, anonymisation, LLM-based scoring, and feedback generation, is available at \url{https://github.com/Integrated-Intelligence-Lab/llm_assisted_grading.git}.

\section{Methods}

To test how LLMs can be effectively incorporated in a real-life grading workflow, we implemented a series of six short in-class ``bonus tests", conducted during the weekly exercise sessions of two undergraduate mathematics courses at the Vrije Universiteit Brussel. Each test lasted 10 minutes and consisted of two handwritten pen-and-paper exercises aligned with the material taught the previous week. Students completed all tests under supervision on a standardised answer sheet designed for automated processing.

 Unlike midterms, which contribute substantially to the final grade, the bonus tests were designed as low-stakes formative checkpoints to encourage continuous engagement early in the semester. Performance across the six tests was rescaled to contribute up to 1.0 additional point on top of the final exam grade scored out of 20. Because the tests were brief, frequent, and structurally uniform, they produced a consistent stream of comparable handwritten responses. This made them well-suited for evaluating the LLM-assisted grading and feedback workflow under realistic classroom conditions.

\paragraph{Student instructions.}
Students received all necessary instructions directly on the test sheet (see Appendix~\ref{a:grading sheet}), including how to complete the required identifiers (ID bubbles, group, and version) and where to write their answers (within the designated answer boxes), without imposing any special writing or legibility instructions beyond standard exam practice. In addition, a pilot bonus test was administered earlier in the course to familiarise students with the sheet format and the grading procedure. By the time of the evaluated tests, students were therefore accustomed to the template and its administrative requirements.

\subsection{Test Sheet Structure and Student Identifiers}

To enable reliable automated processing of handwritten work at scale, the bonus tests were administered on a standardised answer sheet, designed for OCR parsing and downstream LLM-based grading as shown in Appendix~\ref{a:grading sheet}. Each sheet contained two clearly delineated answer boxes, one for each test question, along with a separate region for student identification and test metadata, such as student group.

Student identification was captured using a bubble-coded identifier printed on the answer sheets. Students filled in a row of bubbles corresponding to the digits of their university-issued student number. This choice facilitated communication and reduced administrative overhead, but it also introduced fragility: standard student numbers are not designed to tolerate OCR errors or minor filling mistakes, which in practice necessitated manual verification to ensure correct test assignment.

During post-processing, we bulk-scanned the answer sheets into a single PDF. An OCR and template-recognition step then identified the sheet’s fixed structural elements, including the ID bubbles, the group and version fields, and the two answer boxes. Using these coordinates, each answer box was automatically cropped into a separate image, and all regions containing identifying information were removed prior to LLM submission. This yielded fully anonymised images containing only the student’s handwritten mathematical reasoning. Finally, the OCR-extracted identifiers were manually verified against the recorded student ID.

\subsection{Grading}

Our grading workflow consisted of three components: (1) the construction of solution keys, (2) the development of detailed grading keys used by the LLM, and (3) the LLM-based grading procedure followed by human verification. As both courses are instructed in different languages, we tested whether the language of the solution and the grading key affected the final output in a pilot bonus test, which served both to familiarise students with the system and to streamline our setup. A pilot bonus test was used to assess whether instruction language influenced grading. Agreement between languages was near perfect (quadratically weighted Cohen's $\kappa = 0.97$), indicating no meaningful effect. Consequently, all backend grading was standardised to English across both courses.

\subsubsection{Solution Keys}

For each question, we prepared a fully worked-out LaTeX solution that followed as closely as possible the methods and notation emphasised during lectures and exercise sessions. These solutions provided a reference for the LLM during grading.  

\subsubsection{Grading Keys}

In parallel with the construction of fully worked-out solution keys, we developed detailed grading keys that specify how points are allocated across individual reasoning steps. Among all components of the workflow, the grading key proved to be the most sensitive to LLM interpretation. Large language models tend to follow instructions very literally, and even minor ambiguities or underspecified criteria can lead to inconsistent or unintuitive scoring behaviour. As a result, the grading key required substantially more iterative refinement than other elements of the pipeline.

Through repeated testing and adjustment across the six bonus tests, we identified a set of practical principles for constructing grading keys that align with the principles set out by~\cite{liu2024ai} and yield stable and interpretable LLM grading. First, grading keys must decompose solutions into small, explicitly enumerated steps, each associated with a fixed and relatively low point value (typically 2–3 points out of 10 per step). This fine-grained structure reduces variance by preventing single interpretive decisions from disproportionately affecting the final score. Second, when multiple valid solution paths exist, acceptable alternatives should be explicitly allowed for in the grading key rather than left to implicit model inference. Finally, vague formulations such as ``partial credit'' were avoided in favour of concrete, operational criteria, including explicit statements of which deviations should not result in point deductions. Next to the question-specific grading key, we also provided additional grading instructions in the grading prompt, relevant to the entire course, such as the solution space. See the prompt in Section~\ref{sec:LLM-Based Grading Procedure}. Moreover, the grading prompt instructs the model to assess conceptual understanding by following the student’s logic step by step, rather than focusing solely on the final answer. 

The most recurrent mismatch between the grading keys' intent and the LLM's interpretation was overly literal execution of underspecified criteria. This issue is illustrated in Figure~\ref{fig:grading_key_example}, which shows an early grading key with two concrete design flaws. First, the key assigns a large fraction of the total score to a single intermediate step (``factorisation''), making the final grade overly sensitive to one loosely specified criterion. Second, the term ``factorise'' is imprecise in this context and led the model to award credit whenever any factorisation appeared in the student's work, even when it was mathematically irrelevant to solving the limit. Another recurrent issue is over-optimism: the model actively searches for a correct-looking expression in the student’s work and may assign full credit when it encounters an intermediate result that resembles the final answer, despite the absence of a complete solution. To combat this second issue as much as possible, the grading prompt explicitly instructs the model to ignore steps that do not follow from solid mathematical reasoning.

\begin{figure}[t]
\centering
\begin{tcolorbox}[colframe=mygrey]
\textbf{Grading key example: }\\

Give a score out of 10 for the student's explanation. 
Give 2 points for recognizing that both numerator and denominator are zero at this point. \colorbox{red}{4 points for factoring correctly.} 2 points for cancelling the common part in the numerator and denominator. 2 points for the final solution. If the student uses l’Hôpital instead: 2 points for recognizing that both numerator and denominator are zero at this point. 2 points for recognizing that l'Hôpital can be applied.  4 points for the derivative and 2 points for the final solution. If the student uses a different approach that is still mathematically sound, you can overrule the given grading scheme and assign your own sub-grades; in that case, also flag that the student used a different approach. Finally, on a new line at the very end, write exactly:\\
Total: X/10
(where X is the numeric score).\\
Flag: 0/1 (Where 1 means the student used a different approach)\\
Finally motivate your grade.
\end{tcolorbox}
\caption{
Example grading key highlighting two design issues: assigning too many points to a single step and using imprecise terminology (“factorise”), which can cause the model to award credit for mathematically irrelevant operations.
}
\label{fig:grading_key_example}
\end{figure}

\subsection{LLM-Based Grading Procedure}
\label{sec:LLM-Based Grading Procedure}
The grading itself was carried out using GPT-5.1, operating on the anonymised answer fragments extracted from the scanned test sheets. Each student's response was graded independently five times to account for the inherent stochasticity of LLM outputs. On each pass, the model received the same structured input consisting of: the question text, the fully worked solution, the detailed grading key, and the anonymised cropped image of the student’s handwritten answer.

The prompt used for each evaluation followed a fixed template:

\begin{tcolorbox}[colframe=mygrey]
\textbf{Grading prompt: }\\

You are grading a student's solution. Make sure the question they are attempting to solve matches the question provided below. If it does not match, give them a score of 0.\\

Assume the course works over the real numbers $\mathbb{R}$; do not penalise students for omitting complex solutions or incomplete complex solutions. Award credit for intermediate steps only if they are explicitly written by the student; do not infer non-trivial reasoning from a correct final answer except for trivial algebraic simplifications. Base all scoring strictly on evidence in the student's solution; if evidence is missing, assign zero for that criterion, do not guess or hallucinate it. Be aware of nonsensical segments; apply partial credit only to valid, supported steps, and do not award credit for reasoning that would not imply the subsequent result.\\

Question:
\{question text\}

Correct solution:
\{solution text\}

Grading key:
\{grading key\}

Student answer:
\{anonymised answer image\}
\end{tcolorbox}

We incrementally developed this prompt to ensure that the model grounded its evaluation as much as possible in the official solution and grading key, without deviating into alternative interpretations, solution strategies, or hallucinations.

To derive a provisional score from the five evaluations, we opted for a conservative approach in the student's favour by selecting the maximum of the five LLM-generated scores. This approach protects students from accidental under-scoring, which is typically more detrimental than occasional over-scoring~\cite{brockner1987self, brookhart2012grading}. However, selecting the maximum introduces the possibility of selecting an outlier when the model produces an unusually high grade within a cluster of lower, more consistent ones. To surface such cases, we computed several internal consistency measures across the five scores, including the variance, the minimum–maximum spread, and a simple anomaly score based on deviation from the sample mean. Responses exhibiting substantial disagreement across evaluations were automatically flagged for review. In practice, however, no stable threshold or deviation pattern emerged that would reliably separate harmless stochastic variation from genuine grading errors. This finding indicates that, for the current generation of LLMs, automated inconsistency detection alone is insufficient, and underscores the need for a human-in-the-loop design in which final grading authority remains with human instructors.

Representative graded outputs are provided in Appendix~\ref{Graded Examples} (Figure~\ref{fig:graded_examples}), which shows three synthetic responses to the same question receiving scores of 0/10, 3/10, and 10/10. These examples illustrate how the grading key is operationalised in practice and how the structured LLM explanation supports subsequent human verification.

\subsection{Human Verification and Calibration}

Although the LLM provided an initial grade, final responsibility for all scoring remained with the instructors. Each question was presented to a human grader via a compact PDF report containing:

\begin{itemize}
    \item the anonymised student answer image,
    \item the five LLM-generated scores,
    \item the consistency metrics flag,
    \item and the provisional (maximum) score.
\end{itemize}

The human grader then reviewed each provisional score, along with the LLM's reasoning behind the grade, and either accepted it or overruled it. To ensure that the LLM grade was not an unreliable anchor, a randomly selected subset of responses was graded independently by teaching assistants who were blind to the LLM outputs, see section~\ref{sec: Grading Alignment and Consistency} for more details. 

With this hybrid setup, we aimed to preserve the efficiency benefits of automated grading while ensuring that final decisions remained accurate, fair, and in line with pedagogical intent.

\subsection{Analyses}
\label{sec:analyses}
We report three analyses addressing complementary questions about classroom deployment. First, we quantify time savings by comparing grading time under manual versus digital grading; because graders experienced both media, we counterbalance the order (manual-first vs.\ digital-first) and report timings under this design. Second, we assess grading alignment by comparing human vs. human agreement (manual grading) to human vs. LLM agreement (digital grading) using quadratically weighted Cohen's $\kappa$ per question instance, treating inter-annotator variability as a pragmatic benchmark for acceptable grading variation~\cite{morris2025automated}. Third, to account for LLM stochasticity, we evaluate grading alignment under two aggregation rules over five independent model evaluations, using both the median and the maximum LLM grade as the provisional score.   

\section{Results}

To quantify time savings and grade calibration in our proposed workflow, we compared grading speeds between manually graded tests and tests graded with LLM assistance (hereafter ``digital''), i.e., with GPT's provisional maximum score and accompanying rationale. Six experienced graders participated in the timing and calibration study, each grading two different questions drawn from three bonus tests (six questions in total). For each question instance, two graders (denoted A1 and A2) were drawn from this pool and independently graded the same set of 30 responses under the specified medium/order. Grading was conducted in two rounds, enabling grader pairing and counterbalancing of the order in which the two grading media were experienced.

In order to minimise learning and familiarity effects, the experimental design incorporated several controls. First, graders alternated both the question and the grading medium order between rounds, ensuring that no grader evaluated the same test in both rounds and that the order of manual versus digital grading varied in both rounds for each participant. Second, before timed grading began, each grader completed a warm-up phase in which they graded 15 out-of-sample responses of the same question type. These 15 warm-up questions ensured that the measured time reflected steady-state grading rather than initial grading key familiarisation. Finally, in both conditions, grading was performed digitally on scanned tests to avoid confounding effects from manual transcription or student-identifier lookup; this choice slightly favours manual grading relative to traditional pen-and-paper correction and therefore yields a conservative estimate of speed gains.

\subsection{Grading Time Comparison: Manual vs. LLM-Assisted}

Across graders and questions, we analysed digital-to-manual grading-time ratios on the log scale (i.e., using $\log(\mathrm{D/M})$) and exponentiated to report the geometric mean ratio. Aggregating per question instance under the counterbalanced order design, the geometric-mean D/M ratio was $0.767$ (95\% CI: $[0.595,\,0.989]$), corresponding to an average time reduction of $23.3\%$ (95\% CI: $[1.1\%,\,40.5\%]$) for LLM-assisted grading relative to manual grading. This pattern was observed across graders with different baseline grading speeds, indicating that the effect was not driven by a small subset of fast or slow annotators. In two cases, graders were slower when digital grading was their first grading medium; however, even these graders were faster in their second round when the digital medium came second. Conversely, for all graders, digital grading was faster than manual grading when digital grading occurred second.

Taken together, these results suggest that the grading medium itself has a more consistent impact on grading speed than the order in which grading conditions are experienced. While some initial adaptation costs may arise when graders first encounter the digital workflow, these are outweighed by the structural efficiency gains provided by LLM-assisted grading once the process is underway. 

\begin{table}[t]
\centering
\begin{tabular}{lcc}
\toprule
Question & D/M& D/M \\
\midrule
Bonus 3 -- Q2B & \textbf{0.705} & 1.323 \\
Bonus 4 -- Q2A & \textbf{0.998} & 1.165 \\
Bonus 5 -- Q1  & 0.907 & \textbf{0.696} \\
\midrule
Bonus 3 -- Q2A & 0.596 & \textbf{0.632} \\
Bonus 4 -- Q1  & 0.577 & \textbf{0.675} \\
Bonus 5 -- Q2A & \textbf{0.629} & 0.655 \\
\bottomrule \\
\end{tabular}
\caption{
Paired ratios of digital to manual grading time (D/M) for each question.
Each row corresponds to one question graded by two annotators in opposite order:
one graded manually first (M$\rightarrow$D, shown in \textbf{bold}),
the other digitally first (D$\rightarrow$M).
Ratios below 1 indicate faster digital grading.
}
\label{Timings_table}
\end{table}

\subsection{Grading Alignment and Consistency}
\label{sec: Grading Alignment and Consistency}

In addition to grading speed, we examined how LLM-assisted grading affected alignment and consistency of assigned scores. During the study, several annotators reported feeling more consistent and confident in their grading decisions when using the digital workflow, particularly because the grading key and model-generated reasoning provided a stable reference during evaluation. We therefore assessed whether this subjective experience was reflected in objective agreement measures.

\begin{figure}[ht]
    \centering
    \includegraphics[width=\linewidth]{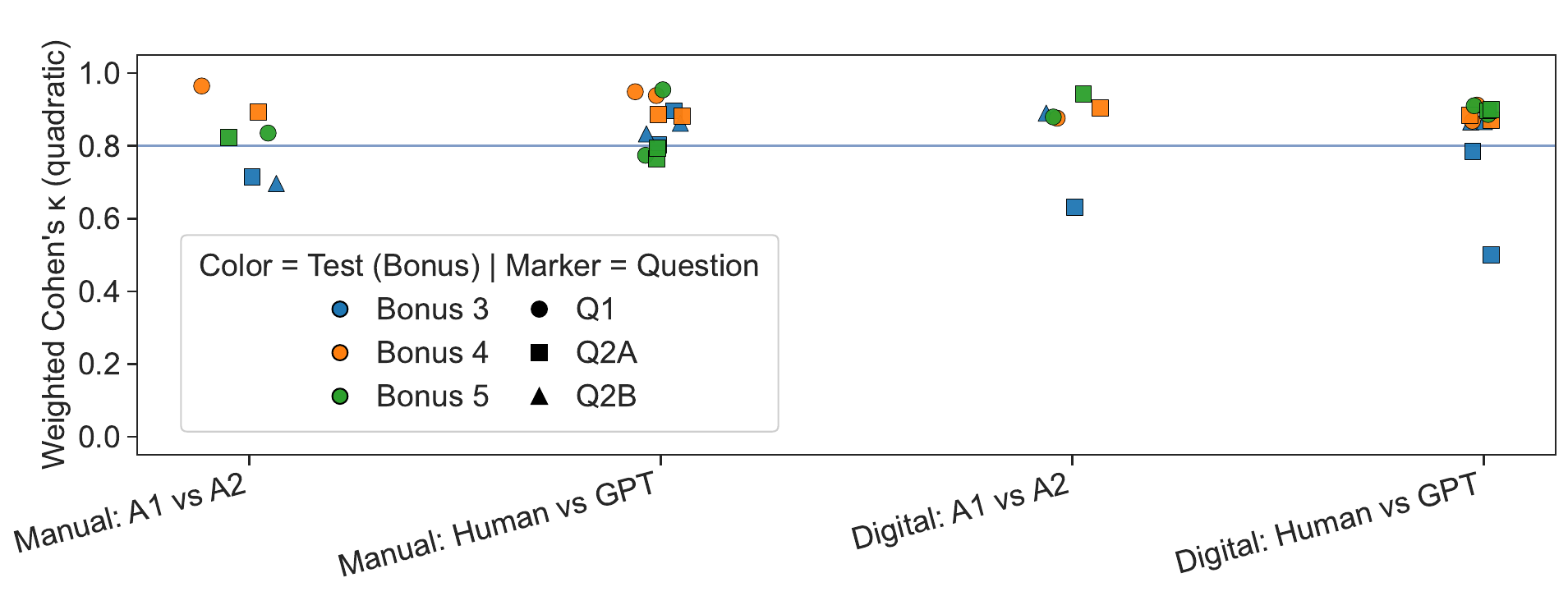}
    \caption{Quadratically weighted Cohen’s $\kappa$ values comparing grading agreement across media, tests, and question types. A1 and A2 denote the two human annotators (e.g., A1 vs A2 indicates inter-annotator agreement). Colours indicate the bonus test, while marker shapes denote the question type. Across all but one question–annotator pairing, annotator–LLM agreement in the digital condition is comparable to, and in several cases higher than, inter-annotator agreement under manual grading.}
    \label{fig:kappa_overview}
\end{figure}

Figure~\ref{fig:kappa_overview} reports quadratically weighted Cohen’s $\kappa$ values by grading medium, test, and question type~\cite{morris2025automated}. Across all six question instances, agreement levels were generally high and consistent with empirical inter-rater agreement reported for handwritten mathematics grading by human assessors~\cite{pantzare2015interrater, moons2025checkbox}. Recent studies evaluating GPT-4 for grading handwritten or open-ended mathematics responses report mixed but generally substantial alignment with human graders when structured rubrics are provided, while also documenting non-negligible error rates and consistency issues that prevent fully automated deployment~\cite{caraeni2024evaluating, gandolfi2025gpt, lee2025can, liu2024ai}. Against this backdrop, we find that agreement between annotators and the LLM (GPT-5.1) in the digital condition is comparable to, and in several cases higher than, the agreement observed between the two human annotators in the manual condition. This pattern is particularly evident for Bonus~3~Q2B and Bonus~5~Q2A, where both annotators individually agreed more closely with the LLM than with each other.

Table~\ref{tab:kappa_summary} reports the corresponding quadratically weighted Cohen’s $\kappa$ values underlying Fig.~\ref{fig:kappa_overview}, shown per question instance. In addition to the primary $\kappa$ values computed using the maximum of five LLM evaluations, the table also reports agreement obtained when using the median LLM score. The two aggregation strategies yield very similar agreement levels across all questions, with median-based aggregation generally producing slightly higher $\kappa$ values. This modest increase could be due to fewer outlier scores when using median aggregation than when the maximum LLM score is used as the provisional grade.

\begin{table}[ht]
\centering
\begin{tabular}{lccc}
\toprule
Question & Manual: A1 vs A2 & Digital: A1 vs GPT & Digital: A2 vs GPT \\
\midrule
Bonus 3 -- Q2A & 0.71 & 0.79 (0.71)  & 0.50 (0.43) \\
Bonus 3 -- Q2B & 0.70 & 0.87 (0.85) & 0.87 (0.85) \\
Bonus 4 -- Q1  & 0.96 & 0.87 (0.94) & 0.91 (0.91) \\
Bonus 4 -- Q2A & 0.89 & 0.89 (0.85) & 0.87 (0.86)\\
Bonus 5 -- Q1  & 0.84 & 0.91 (0.89) & 0.89 (0.93) \\
Bonus 5 -- Q2A & 0.82 & 0.90 (0.88) & 0.90 (0.88) \\
\bottomrule
\end{tabular}
\caption{Quadratically weighted Cohen's $\kappa$ values compare inter-annotator agreement under manual grading with annotator--LLM agreement under digital grading, shown for each question instance. In the teaching workflow, we used the maximum of five independent LLM evaluations as the provisional score (a conservative choice in the student's favour). For analysis, we also report results using the median of five evaluations, which is less sensitive to outliers. In this table, the primary $\kappa$ values (without parentheses) use the median of five LLM evaluations as the provisional grade, while the values in parentheses use the maximum LLM score instead. Both aggregation strategies yield similar agreement levels, with median-based aggregation typically producing slightly higher $\kappa$ values. The low alignment on Bonus~3--Q2A was driven by differing interpretations of the grading key: both annotators applied an additional grading step that was not explicitly specified, underscoring the need for a detailed and comprehensive grading key.
}
\label{tab:kappa_summary}
\end{table}

\begin{figure}[ht]
    \centering
    \includegraphics[width=\linewidth]{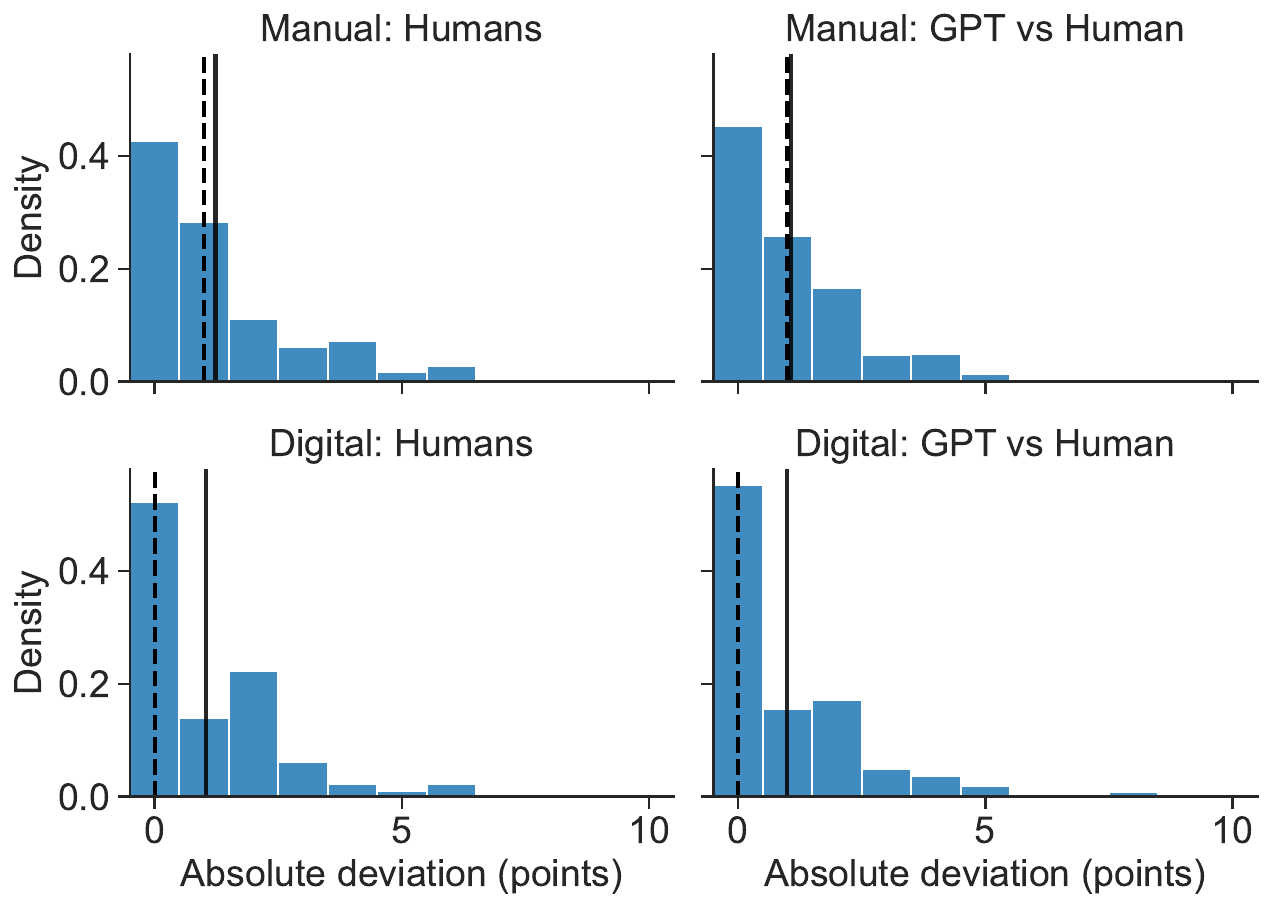}
    \caption{Distributions of absolute score deviations for all grader pairings under manual and digital grading. The top row shows manual grading comparisons (human vs. human, human vs. GPT using the median assigned score), while the bottom row shows digital grading comparisons. Dashed vertical lines indicate median deviations and solid lines indicate mean deviations. In the manual setting, mean and median deviations coincide for both human vs. human and human vs. GPT comparisons, indicating similar grading behaviour. In the digital setting, median deviations are lower than the mean for both human vs. human and human vs. GPT comparisons, consistent with an anchoring effect in which the LLM grade acts as a stabilising reference that reduces typical disagreement while leaving mean deviation largely unchanged.}

    \label{fig:mad_distributions}
\end{figure}

To complement agreement statistics, Figure~\ref{fig:mad_distributions} shows distributions of absolute score deviations with respect to the median grade assigned by GPT-5.1. We focus on the most informative comparisons in each condition: inter-annotator deviations in the manual setting (human–human) and deviations between annotators and the LLM in the digital setting (human vs. GPT). Two patterns stand out. First, in the digital setting, deviations are more tightly concentrated near zero than in the manual setting, reflected in substantially lower median absolute deviation values. At the same time, mean absolute deviation remains comparable across settings, indicating that overall average disagreement does not change markedly. This divergence between mean and median in the digital condition is consistent with an anchoring effect, whereby the LLM grade acts as a stabilising reference that reduces typical disagreement without affecting the average magnitude of deviations. Second, despite this strong central alignment, a small number ($ \approx 3\%$) of larger outliers occur in the human vs. GPT comparisons that do not reach the same magnitude among human annotators. These rare deviations are precisely the cases targeted by the human verification step.

\begin{figure}
    \centering
    \includegraphics[width=1\linewidth]{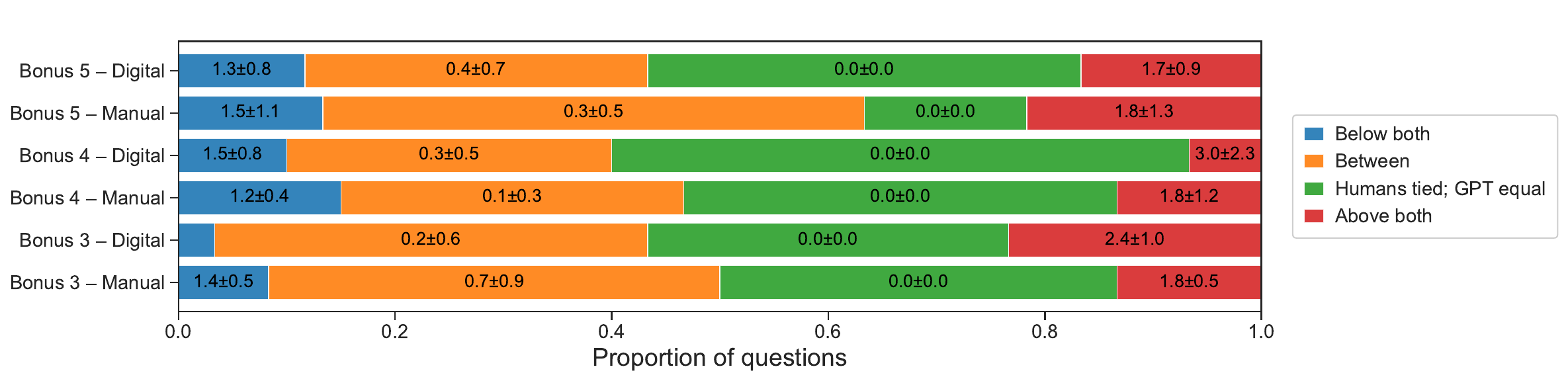} 
    \caption{Relative positioning of LLM-assigned grades with respect to the two human annotators. Bars indicate the proportion of responses for which the LLM score lies below both human scores, between the two human scores, is equal to both, or exceeds both. Error bars report mean $\pm$ standard deviation of absolute distance to the nearest human boundary. Across grading media, the LLM most frequently assigns intermediate or matching scores, with deviations outside the human range occurring less often.}
    \label{fig:gpt_positioning}
\end{figure}

Finally, Figure~\ref{fig:gpt_positioning} examines where the LLM’s median assigned scores fall relative to the two human graders. In both grading media, the LLM most frequently assigns scores that lie between the two human scores or coincide exactly with one of them. To further illustrate this alignment at the level of individual responses, Appendix Figure~\ref{fig:positioning max all dots} plots all five LLM scores alongside the minimum and maximum human grades for each submission. This visualization shows that, for the large majority of cases, the distribution of LLM scores is tightly embedded within the human grading range, and that human scores themselves cluster closely around the LLM median. Moreover, for most submissions, the dispersion of the five LLM evaluations is smaller than the spread between the two human grades, suggesting that the model provides a locally stable reference even when human judgements differ. When the maximum LLM score is used instead of the median, deviations outside the human range become larger in magnitude and predominantly positive, as shown in Appendix Figure~\ref{fig:positioning max}. This difference reflects the aggregation choice rather than a qualitative change in the underlying grading behaviour of the model.

Taken together, the reduction in dispersion and middle-of-the-road grading indicate that, for the large majority of cases, the digital workflow stabilises grading decisions and yields more consistent scores than fully manual grading between human annotators. At the same time, the presence of a small number of large outliers in the human vs. GPT comparisons underscores the need for manual flagging and human verification. These rare cases do not negate the overall consistency gains but instead motivate the hybrid design of the workflow, in which LLM-generated grades act as a stabilising and time-saving reference while final judgement remains with human graders.

\section{Discussion}

We set out to evaluate whether large language models can be integrated into grading and feedback workflows to improve efficiency and consistency without compromising fairness or accuracy. The results show that this is indeed possible, but only when LLMs are embedded within a carefully designed, hybrid workflow that explicitly accounts for their limitations. Among all components of the pipeline, the grading key emerged as the single most critical determinant of reliable performance. We found that LLM-assisted grading is highly sensitive to how evaluative criteria are specified, and that stable, interpretable outputs depend on grading keys that decompose solutions into explicit, low-level steps with unambiguous point allocations.

Another key finding is that LLM-assisted grading yields substantial and robust time savings. Across graders with varying baseline speeds, digital grading was consistently faster than manual grading. Notably, the observed speed-up reflects structural efficiencies rather than transient learning effects. More specifically, once graders engaged with the digital workflow, grading tended to become faster regardless of whether it was encountered first or second.

Beyond efficiency, the results also shed light on grading alignment and consistency. Focusing on the most informative agreement pairs in each condition, human vs. human agreement under manual grading and human vs. GPT agreement under digital grading, we find that LLM-assisted grading achieves alignment that is comparable to, and in several cases tighter than, fully manual grading. In the digital workflow, deviations are more strongly concentrated near zero, reflected in lower median absolute differences, indicating that for most responses the LLM’s grades differ only minimally from those of human annotators.

This pattern holds across all agreement pairs considered. When examining manual grading alone, human annotators disagree with each other to a similar extent as they disagree with the LLM, with a median absolute deviation of one point. However, the shape of the deviation distributions differ. Human–GPT comparisons exhibit a higher density of very small deviations, indicating tighter central alignment, while also showing a small number of larger outliers that are not observed to the same extent in inter-annotator comparisons. Within the digital setting, the presence of a provisional LLM grade and explicit reasoning serves as an effective anchor, increasing alignment among human graders. These quantitative findings align with annotators’ subjective reports of feeling more consistent when grading digitally.

Nevertheless, our analysis underscores that LLMs are not error-free. While most LLM-generated grades closely track human judgment, cases in which the model hallucinates or misinterprets instructions do occur, resulting in a small number of large outliers. These deviations are explicitly accounted for and made visible through multi-pass grading and explicit human verification. Rather than undermining the workflow, these findings reinforce the necessity of its hybrid design: LLMs contribute speed and regularisation, while human graders retain authority over exceptional cases.

Finally, several limitations and directions for future work should be noted. The study was conducted in a limited number of mathematics courses and question types, and results may differ for more open-ended tasks or disciplines where structured grading keys are not feasible. Nevertheless, the workflow itself is not mathematics-specific and could extend to other STEM contexts, including in-class programming assessments, where short, structured tasks require students to demonstrate understanding through incremental steps rather than complete solutions. In-class and on-paper coding evaluations, in particular, may benefit from this approach by preserving authentic demonstrations of mastery. More generally, it remains an open question how well the workflow generalises to settings without tightly defined rubrics. However, LLMs have demonstrated strong performance in combining OCR with contextual understanding, which may outperform traditional large-cohort grading approaches that rely primarily on keyword matching. Unlike keyword-based methods, contextual analysis allows the model to assess whether students not only mention relevant concepts but also apply them correctly and coherently. Exploring this capability in more open-ended assessment formats is a promising direction for future research. Another limitation was the use of student numbers as identification. In rare cases, our OCR step failed to extract these correctly from the bubble sheet, especially when students did not colour the entire bubble or did so with a light colour. In future iterations, student numbers could be replaced with identifiers explicitly designed to maximise pairwise Hamming distance. Such codes function as simple error-correcting identifiers, allowing the intended value to be recovered reliably even when individual bubbles are partially filled, misaligned, or imperfectly detected during OCR processing.

In conclusion, our findings suggest a shift in how LLMs should be positioned within educational assessment. Rather than asking whether LLMs can replace human graders, a more productive question is how they can be used to strengthen human judgement when embedded in workflows that explicitly anticipate and mitigate their failure modes. This study shows that, when designed in this way, LLMs can function as supportive tools that reduce workload, stabilise grading decisions, and enable timely feedback loops at scale, capabilities that are increasingly essential as generative AI reshapes educational practice.

\paragraph{Code availability} The end-to-end pipeline code (processing, anonymisation, grading, and feedback generation) and analysis scripts used in this study are available at \url{https://github.com/Integrated-Intelligence-Lab/llm_assisted_grading.git}

\bibliographystyle{unsrt}  
\bibliography{references}
\newpage
\appendix
\section{Raw Grading Timings}

\begin{table}[ht]
\centering

\begin{tabular}{l l l c c}
\toprule
Bonus test & Question & Annotator & Order & Time (mm:ss) \\
\midrule
3 & Q2B & L  & M$\rightarrow$D & 13:27 \\
3 & Q2B & L  & D$\rightarrow$M & 9:29 \\
3 & Q2B & V  & D$\rightarrow$M & 12:41 \\
3 & Q2B & V  & M$\rightarrow$D & 16:47 \\
\midrule
4 & Q2A & T   & M$\rightarrow$D & 20:53 \\
4 & Q2A & T   & D$\rightarrow$M & 20:51 \\
4 & Q2A & B   & D$\rightarrow$M & 18:32 \\
4 & Q2A & B   & M$\rightarrow$D & 21:36 \\
\midrule
5 & Q1  & A   & D$\rightarrow$M & 18:43 \\
5 & Q1  & A   & M$\rightarrow$D & 16:58 \\
5 & Q1  & M & M$\rightarrow$D & 12:49 \\
5 & Q1  & M & D$\rightarrow$M & 8:55 \\
\midrule
3 & Q2A & T   & D$\rightarrow$M & 19:27 \\
3 & Q2A & T   & M$\rightarrow$D & 11:35 \\
3 & Q2A & A   & M$\rightarrow$D & 16:46 \\
3 & Q2A & A   & D$\rightarrow$M & 10:36 \\
\midrule
4 & Q1  & M & D$\rightarrow$M & 12:20 \\
4 & Q1  & M & M$\rightarrow$D & 7:07 \\
4 & Q1  & V  & M$\rightarrow$D & 12:43 \\
4 & Q1  & V  & D$\rightarrow$M & 8:35 \\
\midrule
5 & Q2A & B   & M$\rightarrow$D & 21:25 \\
5 & Q2A & B   & D$\rightarrow$M & 13:28 \\
5 & Q2A & L  & D$\rightarrow$M & 14:06 \\
5 & Q2A & L  & M$\rightarrow$D & 9:14 \\
\bottomrule\\
\end{tabular}
\caption{
Raw grading times for manual and LLM-assisted digital grading.
Each row corresponds to a single annotator grading 30 responses for one question.
The grading order of the mediums is by D$\rightarrow$M and M$\rightarrow$D whether manual grading (M) or digital grading (D) was performed first.
}
\label{tab:raw_timings}
\end{table}
\newpage
\section{Grading Deviations w.r.t. Max Score}

\begin{figure}[ht]
    \centering
    \includegraphics[width=1\linewidth]{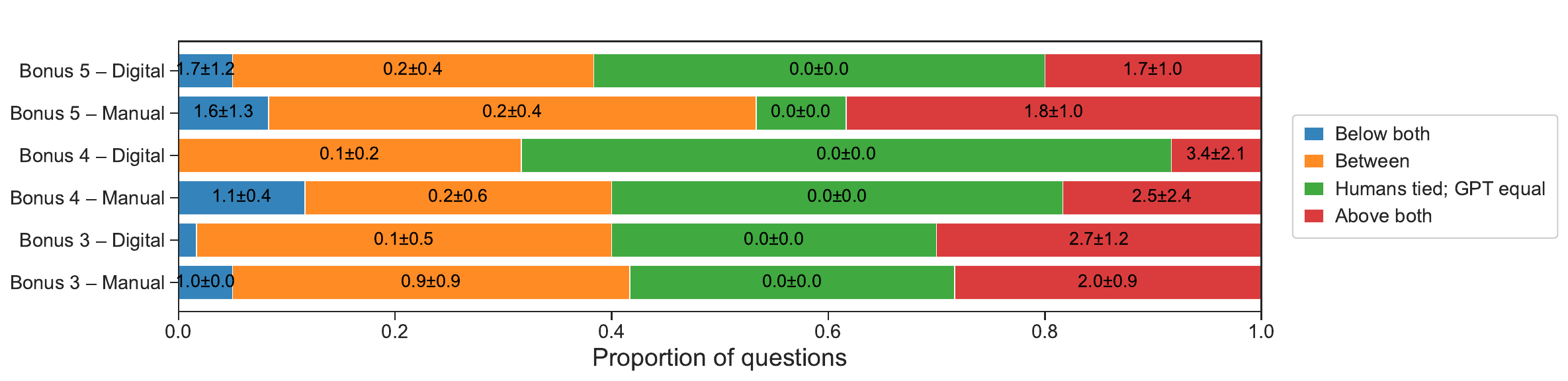}
    \caption{Relative positioning of LLM-assigned grades with respect to the two human annotators when the maximum of five LLM evaluations is used as the provisional grade. Compared to median aggregation (main text), maximum-based aggregation produces more positive deviations outside the human score range.}

    \label{fig:positioning max}
\end{figure}

\begin{sidewaysfigure}
    \centering
    \includegraphics[width=1\linewidth]{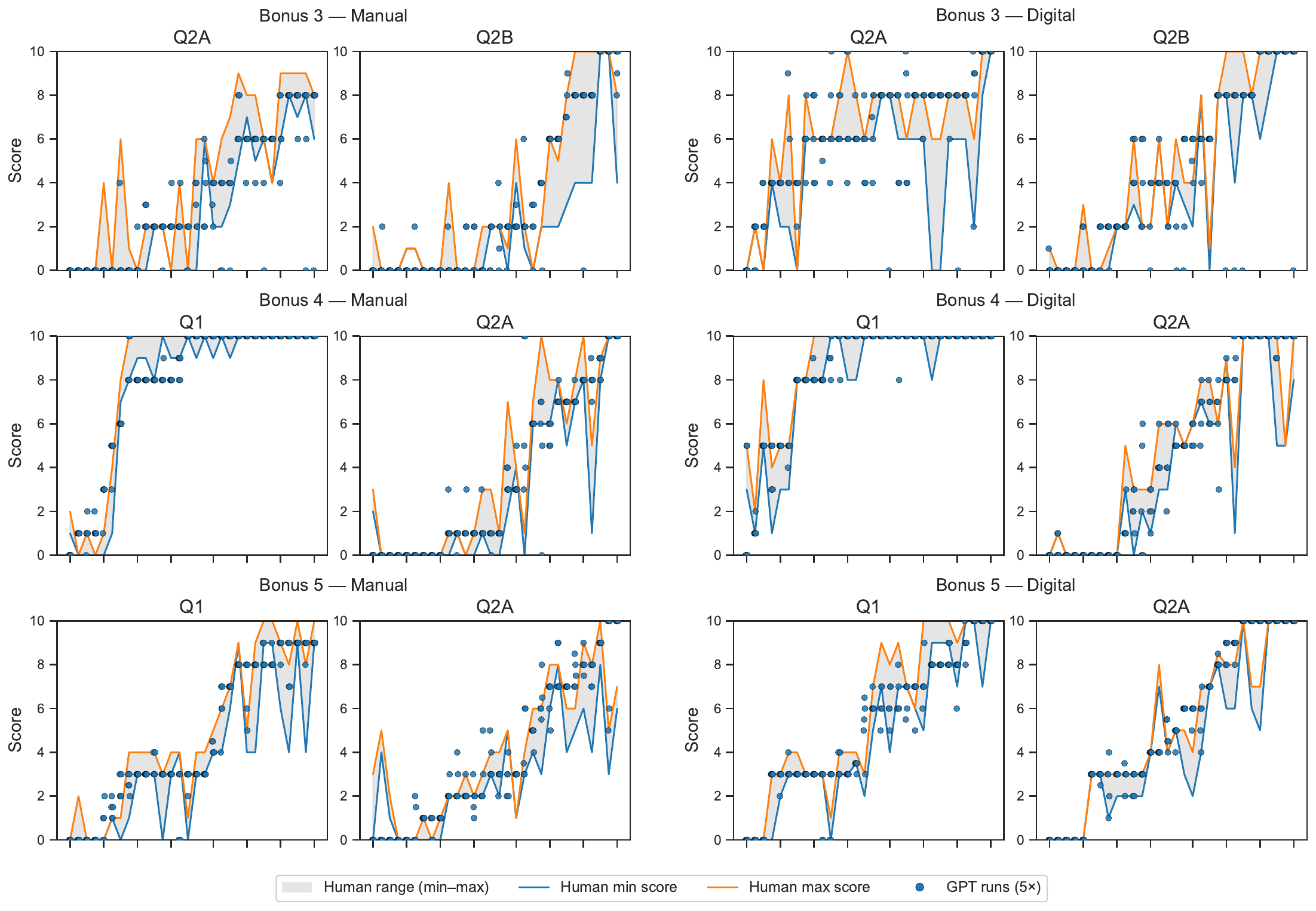}
    \caption{Ordered comparison of five LLM scores with the minimum and maximum human grades per submission, illustrating tight LLM score concentration within the human grading range.}
    \label{fig:positioning max all dots}
\end{sidewaysfigure}

\newpage

\section{Grading Sheet}
\label{a:grading sheet}
\begin{figure}[ht]
    \centering
    \includegraphics[width=0.75\linewidth]{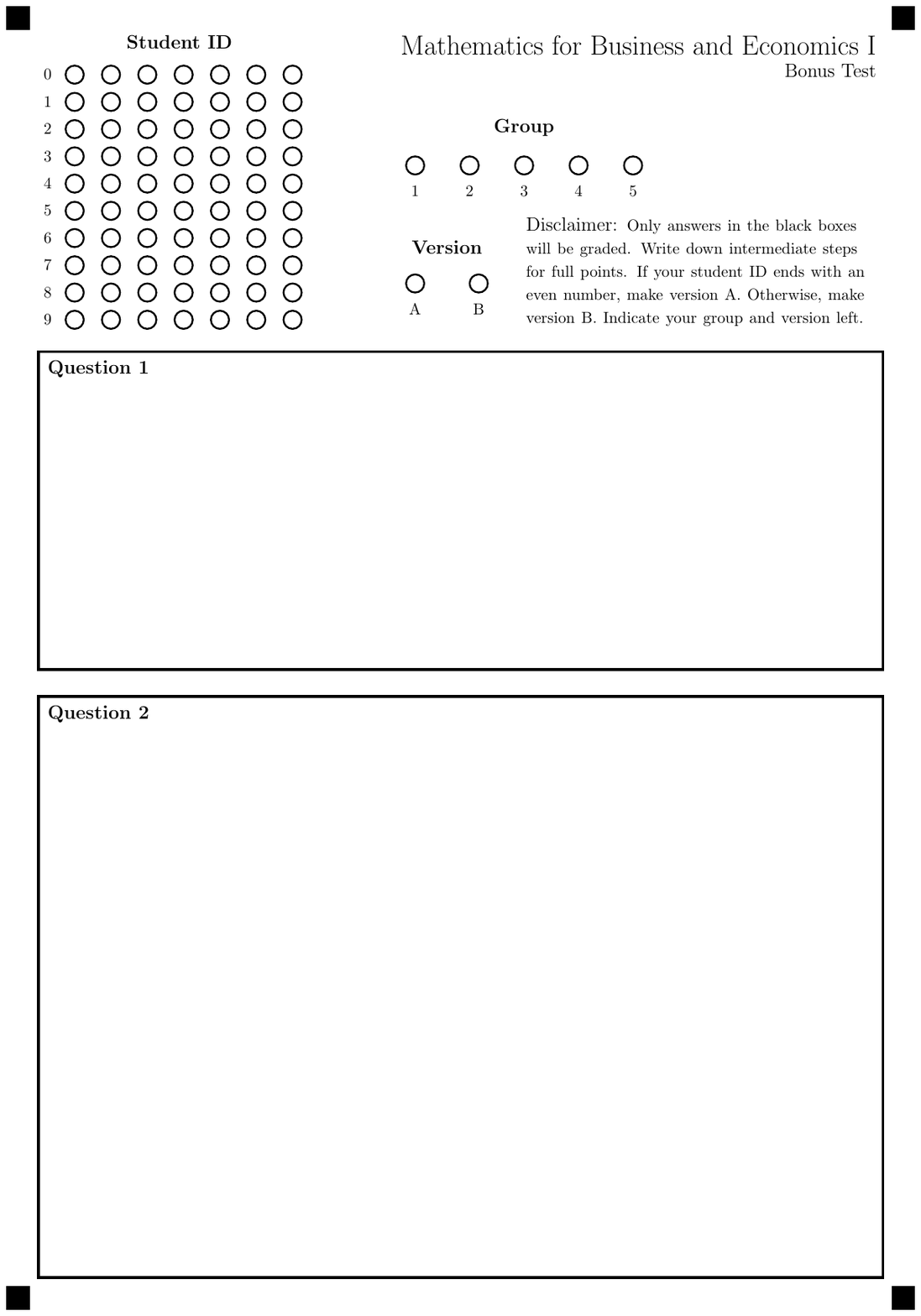}
    \caption{OCR template used for in-class bonus tests. The structured layout enables reliable extraction of student identifiers and automated isolation of answer regions, ensuring that only anonymised handwritten responses are passed to the LLM for grading.}
    \label{fig:OCR template}
\end{figure}

\section{Graded Examples}
\label{Graded Examples}
\begin{figure}
    \centering
    \includegraphics[width=0.85\linewidth]{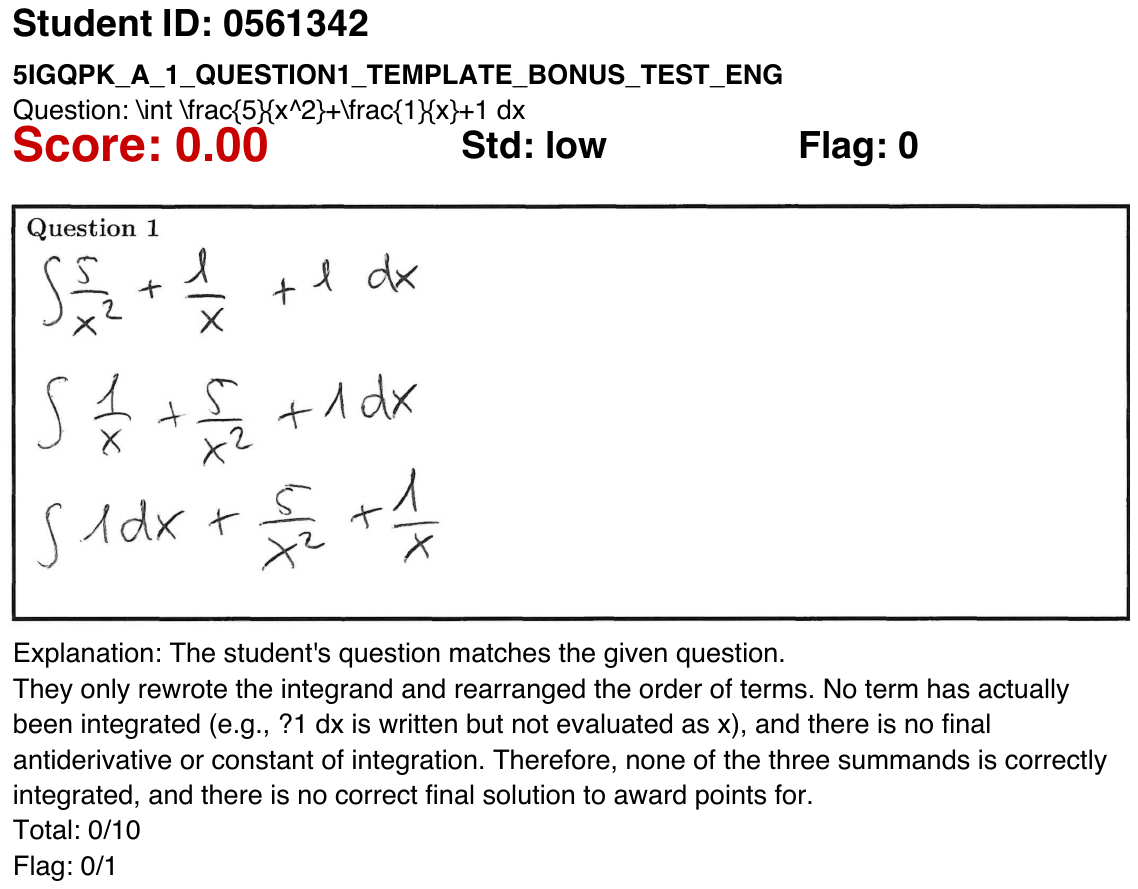}
    \label{fig:Example 1}
\end{figure}

\begin{figure}
    \centering
    \includegraphics[width=0.85\linewidth]{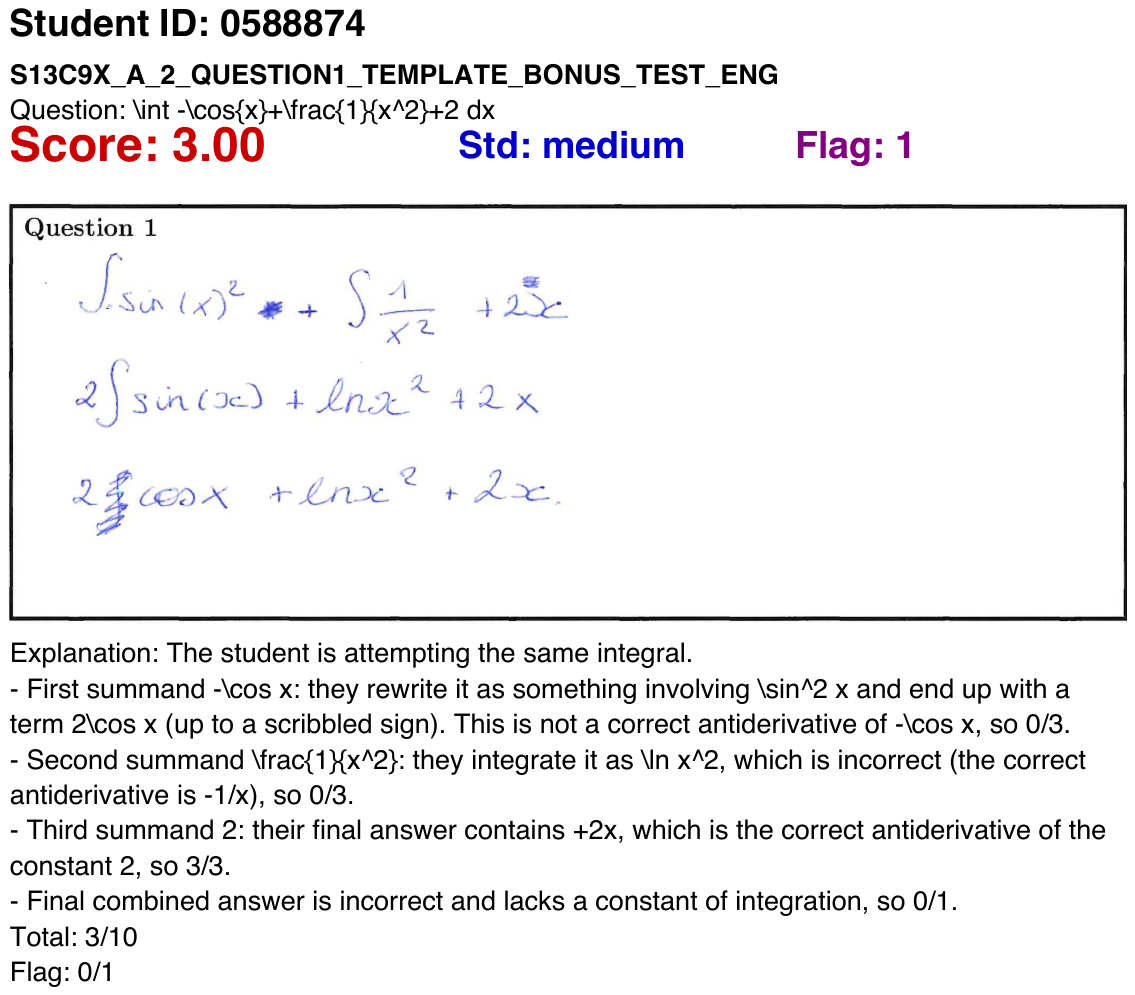}
    \label{fig:Example 2}
\end{figure}

\begin{figure}
    \centering
    \includegraphics[width=0.85\linewidth]{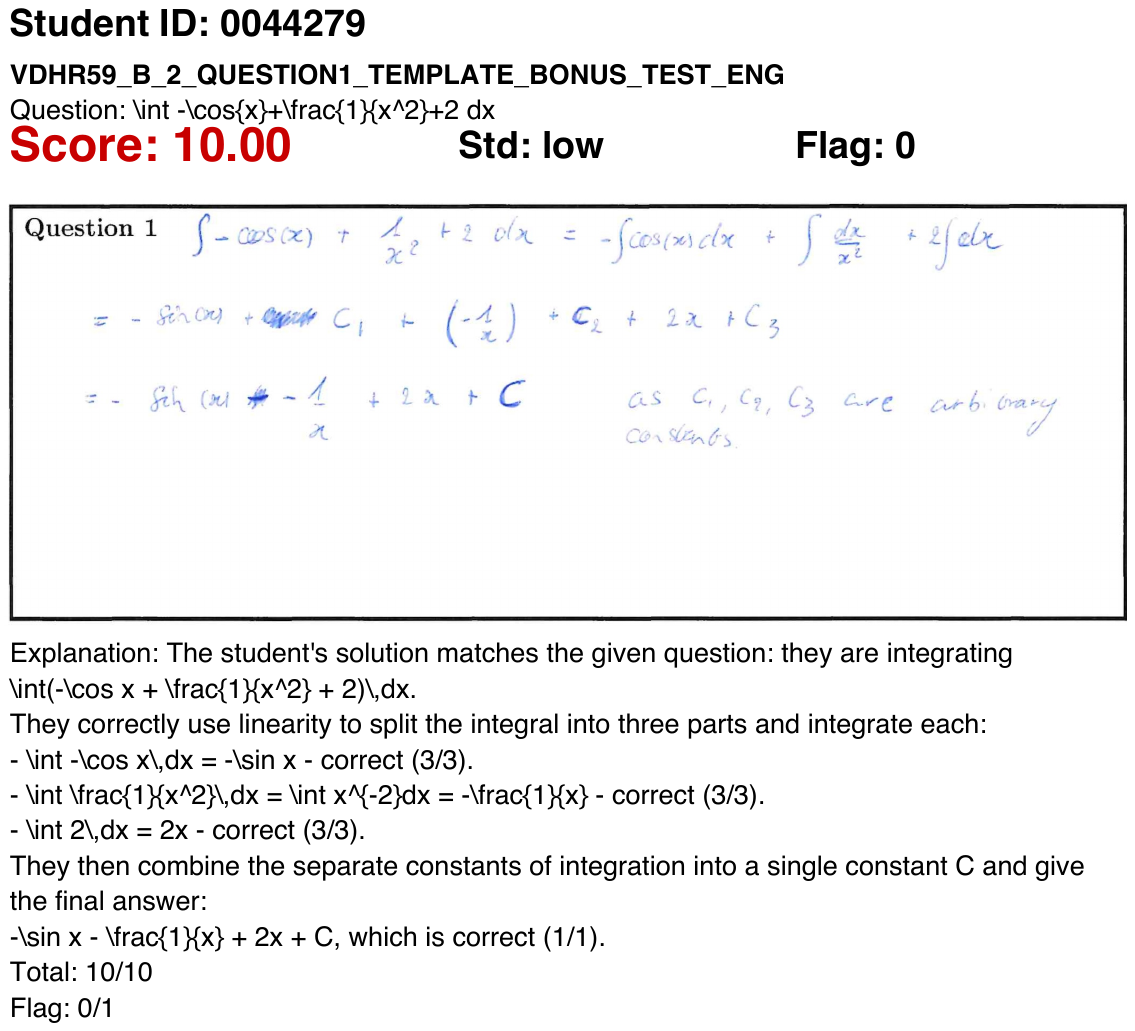}
    \caption{Illustrative graded examples for the same bonus-test question, showing three performance levels: incorrect (0/10), partially correct (3/10), and fully correct (10/10). All responses and student identifiers are synthetic and were created by the authors for demonstration purposes. In each case, the structured LLM-generated explanation follows the grading key and makes explicit how points are allocated across intermediate reasoning steps prior to human verification.}
    \label{fig:graded_examples}
\end{figure}

\end{document}